\documentclass[aps,pre,twocolumn,amsmath,amssymb,showpacs,showkeys]{revtex4}
\usepackage{graphicx}% Include figure files
\usepackage{dcolumn}% Align table columns on decimal point
\usepackage{bm}% bold math

\def\LA{\left\langle}
\def\RA{\right\rangle}
\def\LP{\left(}
\def\RP{\right)}

\begin{document}
\title{A statistical model with a standard Gamma distribution}

%\author{Marco Patriarca}
%%\email{marco@lce.hut.fi}
%\affiliation{Complex Systems Group,
%Laboratory of Computational Engineering,\\
%Helsinki University of Technology,
%P.O.Box 9203, 02015 HUT, Finland}
%
%\author{Anirban Chakraborti}
%%\email{anirban@bnl.gov}
%\affiliation{Department of Physics, Brookhaven National
%Laboratory, \\
%Upton, New York 11973, USA}
%
%\author{Kimmo Kaski}
%%\email{kimmo.kaski@hut.fi}
%\affiliation{%Complex Systems Group,
%Laboratory of Computational Engineering,
%Helsinki University of Technology,
%P.O.Box 9203, 02015 HUT, Finland}

\author{Marco Patriarca}
  \email{marco@lce.hut.fi}
  \homepage{http://www.lce.hut.fi/~marco/}
\author{Kimmo Kaski}
  \email{kimmo.kaski@hut.fi}
  \homepage{http://www.lce.hut.fi/~kaski/}
  \affiliation{Laboratory of Computational Engineering, 
Helsinki University of Technology, P.O.Box 9203, 02015 HUT, Finland
}

\author{Anirban Chakraborti}
  \email{anirban@bnl.gov}
  \homepage{http://www.cmth.bnl.gov/~anirban/}
  \affiliation{
Department of Physics, Brookhaven National Laboratory,
Upton, New York 11973, USA
}

\date{\today}

\begin{abstract}
We study a statistical model consisting of $N$ basic units 
which interact with each other by exchanging a physical entity, according to a 
given microscopic random law, depending on a parameter $\lambda$. 
We focus on the equilibrium or stationary 
distribution of the entity exchanged and verify through numerical fitting of
the simulation data that the final form of the equilibrium distribution is 
that of a standard Gamma distribution.
The model can be interpreted as a simple closed economy in which economic 
agents trade money and a saving criterion is fixed by the saving propensity 
$\lambda$.
Alternatively, from the nature of the equilibrium distribution,
we show that the model can also be interpreted as a perfect gas 
at an effective temperature $T(\lambda)$, where particles exchange
energy in a space with an effective dimension $D(\lambda)$.

\end{abstract}



\maketitle

%%%%%%%%%%%%%%%%%%%%%%%%%%%%%%%%%%%%%%%%%%%%%%%%%%%%%%%%%%%%%%%%%
\section{Introduction}

Statistical physicists like to understand how and why systems evolve from an 
initial toward an equilibrium macroscopic state.
The equilibrium state, far from being just the 
``final state'' of the dynamical evolution, actually reflects
the details of the underlying dynamics.
To write down the ``microscopic equation'' 
governing the dynamics of the evolution is a major goal.
The various probability distributions, resulting from the different
corresponding microscopic equations, have a relevant interest, in that
they can be used to derive most of the macroscopic properties of the system. 
One of the foremost examples is the Maxwell-Boltzmann distribution
for the velocities, which can be obtained as a solution of the
equation which Boltzmann proposed for the evolution of the probability
distribution for a dilute gas.

One of the current challenges is writing down the 
``microscopic equation'' which would correspond to the century old Pareto law 
\cite{Pareto:1897} in Economics,
which states that the higher end of the distribution of income $f(x)$ follows 
a power-law 
$$f(x) \propto x^{-1-\alpha}, \label{pareto}$$
where $x$ is the income (money) and the exponent $\alpha$ has a value
in the interval $1$ to $2$  
\cite{Levy:97,Dragulescu:2001a,Reed:2002,Aoyama:2003}.
In the recent past, several 
studies have been made to investigate the characteristics 
of the real income distribution and provide explanations (see Ref. 
\cite{Slanina:2003} for a brief summary and more references). 

It is our general aim to study a statistical model of closed economy which can 
be either solved exactly or simulated numerically, analyze the microscopic 
equation and investigate what kind of macroscopic money distribution results 
\cite{Chakraborti:2000a,Dragulescu:2000a,Chakraborti:2002a,Hayes:2002a,Chatterjee:2003a,Das:2003}. It would be of particular interest to see whether the study
would give a clue as to why the Pareto law arises. 

In this paper, we study a statistical model consisting of $N$ basic units 
which interact with each other by exchanging a physical entity $x$, according 
to a given microscopic law with one constant parameter $\lambda$. 
We study the stationary probability distributions $f(x)$ for different values
of the parameter $\lambda$. 
We verify through numerical studies that the final form of
the equilibrium distribution $f(x)$ is that of a standard Gamma distribution. 
In Section II, we interpret the model as a simple closed economy 
in which economic agents trade money and a saving criterion is fixed 
by the saving propensity $\lambda$. 
In Section III, using the nature of the equilibrium distribution,
we show that the model can also be interpreted as a perfect gas 
at an effective temperature $T(\lambda)$, made up of 
particles exchanging energy in a space with an effective dimension 
$D(\lambda)$. 
Finally in Section IV, we draw conclusions.

%%%%%%%%%%%%%%%%%%%%%%%%%%%%%%%%%%%%%%%%%%%%%%%%%%%%%%%%%%%%%%%%%
\section{ The Model Economy}

We begin by considering a simple model of closed economy,
in which $N$ agents can exchange money in pairs between themselves.
All the agents can be initially assigned the same money amount $\bar{x}$,
since this condition is not restrictive.
Agents are then let to interact and, at every ``time step'', 
a pair $(i,j)$ is randomly chosen and the transaction carried out.
During the transaction, the agent money amounts $x_i$ and $ x_j$ 
undergo a variation, in which they are randomly reassigned 
between the two agents. The exchange law 
is such that the money is conserved during the transaction, i.\,e.
$x_i + x_j = x_i' + x_j'$, where $x_i'$ and $x_j'$ are the money values after 
the transaction.
This implies that at any time the average initial money $\bar{x}$ 
also represents the average money, $\LA x \RA \equiv \bar{x}$ .

\subsection{Case with $\lambda=0$}
\label{lambda0}
The exchanges are made according to the following law: 
\begin{eqnarray}
  x_i' &=& \epsilon \, (x_i + x_j) \ ,
  \nonumber \\
  x_j' &=& (1-\epsilon) (x_i + x_j) \ ,
  \label{basic}
\end{eqnarray}
where $\epsilon$ is a random number uniformly distributed in 
the interval $(0,1)$.
It can be noticed that, in this model, 
agents have no debts after the transaction, 
i.\,e. they are always left with a money amount $x \ge 0$ or, equivalently,
that $x_i$ is a positive definite quantity, if $\bar{x}>0$.

It can be shown that, as a consequence of the conservation of money,
the system relaxes toward a Gibbs
money distribution~\cite{Chakraborti:2000a,Dragulescu:2000a,Chakraborti:2002a},
\begin{equation}
  f_1(x) = \frac{1}{\LA x \RA} \exp \LP - \frac{x}{\LA x \RA}\RP \ ,
  \label{Gibbs}
\end{equation}
where $\LA x \RA$ represents the average money.
This means that, after the relaxation, most of the agents have 
a very small amount of money, while the number of very rich agents 
is exponentially small.
In other words, for a given $x'>0$, the number of agents with $x > x'$, 
as well as the total amount of money they own, 
exponentially decreases with $x'$.
The equilibrium state represented by the Gibbs distribution (\ref{Gibbs}) 
has been shown to be robust, in that it is reached independently 
of the initial conditions and also in models with multi-agent transactions.

We have re-obtained the exact Gibbs solution for the case $\lambda=0$
by numerical simulations of a system with $N=500$ agents, each agent having 
initially a money amount $\bar{x}=1$.
The system was evolved for $10^6$ time steps -- i.\,e. transactions --
in order to reach equilibrium, and the final equilibrium distributions 
was averaged over $10^5$ different runs.
Figure~\ref{fig1} shows that the numerical results 
(open circles for $\lambda=0$)
are in good agreement with the Gibbs distribution (continuous line).

%%%%%%%%%%%%%%%%%%%%%%%%%%%%%%%%%%%%%%%%%%%%%%%%%%%%%%%%%%%%%%%%%
\subsection{Case with $\lambda>0$}
\label{lambdaNot0}

We now introduce a saving criterion in the same model economy,
by assuming that the saving propensity, which represents the fraction 
of money saved before carrying out the transaction, 
is non-zero, i.\,e. $\lambda > 0$~\cite{Chakraborti:2000a,Chakraborti:2002a}.
Conservation of money still holds, $x_i + x_j = x_i' + x_j'$,
but the money which can be re-assigned in a transaction between
the $i$-th and the $j$-th agent has now decreased by a factor $(1-\lambda)$.
The new law, which replaces Eqs.~(\ref{basic}), is
\begin{eqnarray}
  x_i' &=& \lambda x_i + \epsilon (1-\lambda) (x_i + x_j) \ ,
  \nonumber \\
  x_j' &=& \lambda x_j + (1-\epsilon) (1-\lambda) (x_i + x_j) \ .
  \label{sp1}
\end{eqnarray}
These equations can also be rewritten as follows,
\begin{eqnarray}
  x_i' &=& x_i + \Delta x \ ,
  \nonumber \\
  x_j' &=& x_j -  \Delta x \ ,
  \nonumber \\
  \Delta x &=& (1-\lambda) [\epsilon x_j - (1-\epsilon) x_i] \ ,
  \label{sp2}
\end{eqnarray}
in which money conservation is manifest.

\begin{figure}[ht]
\centering
\includegraphics[width=3.2in]{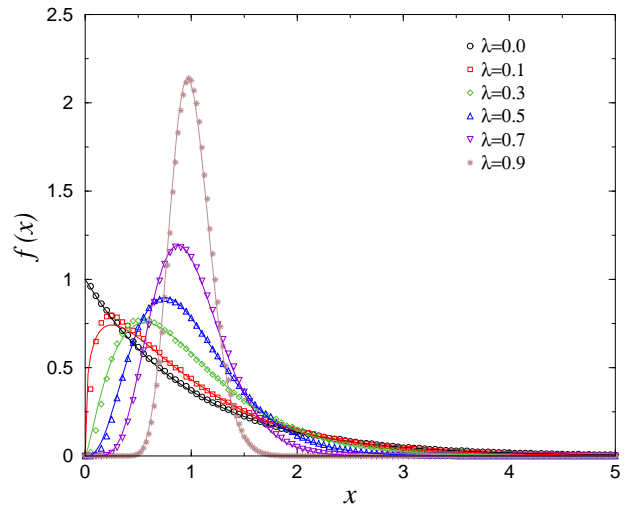}
\caption{
Equilibrium money distributions for different values
of the saving propensity $\lambda$,
in the closed economy model defined by Eqs.~(\ref{basic}) and (\ref{sp2}).
The continuous curves are the fitting functions, defined in Eq.~(\ref{NGibbs})
with the values of $n$ governed by Eq.~(\ref{n}). Note that for the simulation
$\LA x \RA=1$.
}
\label{fig1}
\end{figure}

We studied the equilibrium distribution of this model through
numerical simulations, for various values of $\lambda$,
for $N=500$ agents, again each agent having money $\bar{x}=1$
in the initial state.
In each simulation a sufficient number of transactions,
as far as $10^7$, depending on the value of $\lambda$,
was used in order to reach equilibrium.
The final equilibrium distributions, for a given $\lambda$,
were obtained by averaging over $10^5$ different runs.
The numerical data are shown in Fig.~\ref{fig1} (cases $\lambda\ne 0$).
%_________________________________________
%_________________________________________
%_________________________________________
\begin{figure}[ht]
\centering
\includegraphics[width=3.2in]{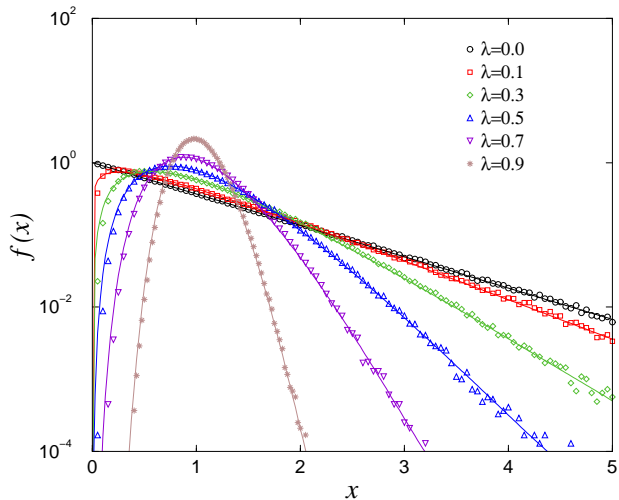}
\caption{
Same quantities as in Fig.~(\ref{fig1}), but on a linear-logarithmic scale.
Note that for the simulation $\LA x \RA=1$.
}
\label{fig2}
\end{figure}
%_________________________________________

We found an analytic form for the equilibrium distribution, 
for a given $\lambda$ ($0 < \lambda < 1$),
which turns out to fit extremely well all data ~\cite{Patriarca:2003}.
The function is conveniently expressed in terms of the parameter
\begin{equation}
  n(\lambda) = 1 + \frac{3 \lambda}{1 - \lambda} \ .
  \label{n}
\end{equation}
This particular form of $n(\lambda)$ was suggested by a mechanical analogy, 
discussed in Sec.~\ref{analogy}, between the closed economy model 
with $N$ agents and the dynamics of a gas of $N$ interacting particles.
Then the money distributions, for arbitrary values of $\lambda$,
are well fitted by the function
\begin{eqnarray}
  f_n(x) &=& a_n x^{n-1} \exp\LP - n x/\LA x \RA \RP \ ,
  \nonumber \\
  a_n    &=& \frac{1}{\Gamma(n)} \left( \frac{n}{\LA x \RA} \right)^n\ ,
  \label{NGibbs}
\end{eqnarray}
where $n$ is defined in Eq.~(\ref{n}) and the prefactor $a_n$,
where $\Gamma(n)$ is the Gamma function,
is fixed by the normalization condition $\int_{0}^{\infty} dx f_n(x) = 1$.

\begin{figure}[ht]
\centering
\includegraphics[width=2.5in]{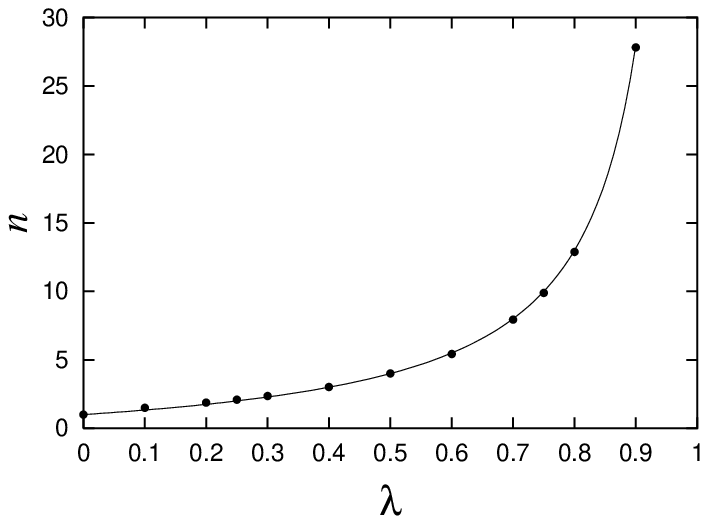}
\includegraphics[width=2.5in]{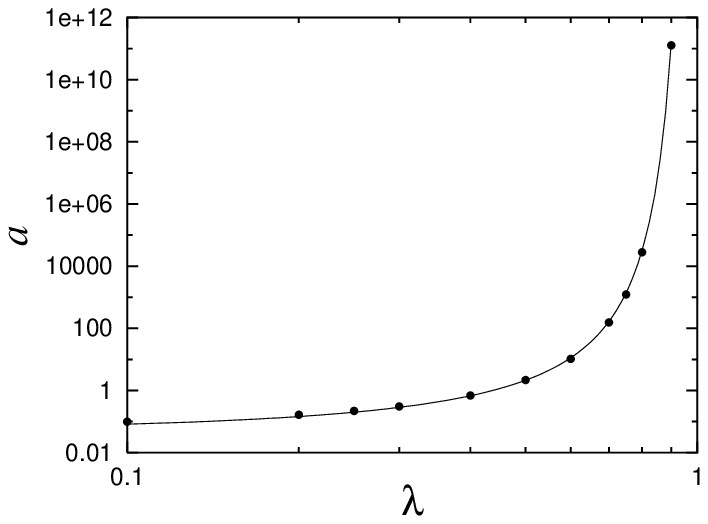}
\caption{
The parameters $n$ (top) and $a_n$ (bottom) versus $\lambda$,
obtained from numerical data (dots) and the corresponding
analytical formulas (continuous curves) 
given by Eqs.~(\ref{n}) and (\ref{NGibbs}), respectively. 
Note that for the simulation $\LA x \RA=1$.
}
\label{figpar}
\end{figure}

The fitting curves for the distribution (continuous curves) 
are compared with the numerical data in Fig.~\ref{fig1}.
The fitting describes the distribution also at large values 
of $x$, as shown by the linear-logarithmic plots in Fig.~\ref{fig2}. In 
Fig.~\ref{figpar}, the numerical values of the parameters $n(\lambda)$ and 
$a_n(\lambda)$ obtained directly from fitting the data (shown as dots) are 
compared with the respective fitting functions (\ref{n}) and (\ref{NGibbs}) 
(shown as continuous curves).

%
%_________________________________________
%_________________________________________
%
%_________________________________________
\begin{figure}[ht]
\centering
\includegraphics[width=2.5in]{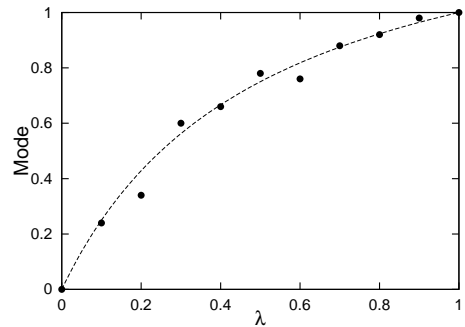}
\caption{
Variation of the mode with the parameter $\lambda$. Dots represent the 
numerical data, while the dashed curve was obtained
theoretically.
}
\label{figmode}
\end{figure}
%_________________________________________
%

By introducing the rescaled variable
\begin{eqnarray}
  \xi         &=& n \, x / \LA x \RA \ ,
  \label{xn}
\end{eqnarray}
the probability distribution (\ref{NGibbs}) can be rewritten as
\begin{equation}
  \frac{\LA x \RA}{n} \, f_n(x) =      
           \frac{1}{\Gamma(n)} \, \xi^{n-1} \exp \LP - \xi \RP 
           \equiv \gamma_n(\xi) \ ,
  \label{gamma}
\end{equation}
where $\gamma_n(\xi)$ is the standard Gamma 
distribution~\cite{Ross:1970,Eric:1999}.
The cumulative distribution for $\gamma_n(\xi)$ is the incomplete
Gamma function $\Gamma(\xi,n) = \int_{\xi}^\infty d\xi' \, \gamma_n(\xi')$,
\begin{equation}
  \gamma_n(\xi) \equiv - \frac{d}{d\xi} \frac{\Gamma(\xi,n)}{\Gamma(n)} \ .
  \label{Gamma}
\end{equation}
The Gibbs distribution (\ref{Gibbs}) is a special case for~$n=1$. 
The term $\LA x \RA/n$, on the LHS of Eq.~(\ref{gamma}), is just
the scaling factor appearing when the change of variable, 
from $\xi$ to $x$, is made in the last equation above
in order to obtain the distribution $f_n(x)$ for the variable $x$.

We notice that, with respect to the Gibbs distribution 
(\ref{Gibbs}), the distribution defined by Eqs.~(\ref{NGibbs}) 
contains the power $x^{n-1}$ and the factor $n$ in the exponential, 
which qualitatively change the distribution shape.
First, they lead to a mode, $x_m$, different from zero.
The mode is shown as a function of the parameter $\lambda$ 
in Fig.~\ref{figmode}, 
where the dashed curve represents the theoretical prediction that the mode
$x_m = 3\lambda/\LP1+2\lambda\RP$ obtainable from Eq.~(\ref{NGibbs}).
Secondly, the presence of the factor $n$ is relevant for the mechanical 
analogy, considered in detail in section~\ref{analogy}.
Finally, in the limit $\lambda \to 1$ (i.\,e. $n \to \infty$), 
the distribution $f_n(x)$ tends to a Dirac $\delta$-function,
peaked around the average value $\LA x \RA$.
A qualitative picture of the evolution of the shapes of the $f_n(x)$'s,
for $\lambda$ going from zero to unity, is obtained by inspection
of the various curves in Fig.~\ref{fig1}.
A more rigorous derivation of the asymptotic distribution for $\lambda \to 1$
can be made by studying the characteristic function 
$\phi(q)$~\cite{Ross:1970, Eric:1999}.
%Since the asymptotic limit $\lambda \to 1$ is equivalent to letting 
%$n(\lambda) \to \infty$, see Eq.~(\ref{n}), for the sake of simplicity,
%in view of the infinite value to be assumed by $n$, 
%it is equivalent -- and convenient -- to consider only integer values of $n$.
The Gamma distribution $\gamma_1(\xi)$ for the dimensionless variable
$\xi$ has a characteristic function
$\phi_1(q)=\int_{0}^{+\infty}d\xi \,\exp(iq\xi)\gamma_1(\xi)=1/(1 - i q)$.
The characteristic function of the Gibbs distribution
$f_1(x)$ in Eq.~(\ref{Gibbs}) is obtained by rescaling $q$ 
by the constant factor $x/\xi=\LA x \RA$,
\begin{equation}
  \phi_1(q) 
     = \LP 1 - i q \LA x \RA \RP^{-1} \ .
  \label{phi1}
\end{equation}
The characteristic function of the generic Gamma distribution
$\gamma_n(\xi)$ is simply given by the $n$-th power 
of $\phi_1(q)$~\cite{Ross:1970, Eric:1999}, 
$\phi_n(q) = 1/(1 - i q)^n$.
Analogously, the corresponding characteristic function of $f_n(x)$ 
Eq.~(\ref{NGibbs}), is obtained by scaling $q$ by $x/\xi=\LA x \RA/n$,
\begin{equation}
  \phi_n(q) 
      = \LP 1 - i q \LA x \RA/n \RP^{-n} \ .
  \label{phin}
\end{equation}
Thus, in the limit $n\to\infty$ ($\lambda \to 1$), one obtains
\begin{equation}
  \phi_n(q) \to \exp\LP i \, q\LA x \RA \RP \ .
  \label{phi_gen}
\end{equation}
The corresponding distribution is obtained by transforming back the
characteristic function, i.\,e.
\begin{equation}
  f_n(x) = (2\pi)^{-1}\int_{-\infty}^{+\infty} dq \, \exp(-iqx) \phi_n(q) 
         \to \delta(x-\LA x \RA) \ .
  \label{delta}
\end{equation}
This limit shows that a large saving criterion leads to a final state
in which economic agents tend to have the same amount of money and,
in the limit of $\lambda \to 1$, they all have the same amount 
$\LA x \RA$.

%%%%%%%%%%%%%%%%%%%%%%%%%%%%%%%%%%%%%%%%%%%%%%%%%%%%%%%%%%%%%%%%%%%%
\section{ The Gas Model}
\label{analogy}

The equilibrium distributions (\ref{Gibbs}) can also be
interpreted as the Gibbs distribution of the energy $x$, 
for a gas at temperature $T = \LA x \RA/k_B$.
This establishes a link between the type of closed economy models 
considered here and statistical systems, suggesting 
a re-interpretation of the economy model 
in terms of a mechanical system of interacting particles.
The introduction of a saving parameter $\lambda>0$ changes the shape of 
the Gibbs distribution into that of a Gamma distribution, 
but the correspondence with a mechanical system is lost only apparently.
In fact, the Gibbs distribution (\ref{Gibbs}) can represent 
the distribution of kinetic energy $x$ only
in $D=2$ dimensions, when its average value is given by 
$\LA x \RA_2 = 2(k_B T/2)$.
In all other cases ($D \ne 2$), it is easy to show, starting from the
Maxwell-Boltzmann distribution for the velocity in $D$ dimensions, 
that the equilibrium kinetic energy distribution $f(x)$ coincides,
apart from a normalization factor, with the Gamma-distribution 
$\gamma_n(\xi)$ with $n=D/2$ for the reduced variable 
$\xi=Dx/2\LA x \RA_D$,
\begin{eqnarray}
  f(x) &=& 
  \frac{ \left( \frac{D}{2\LA x \RA_D} \right)^{D/2} }{\Gamma(\frac{D}{2})}
  x^{D/2-1}\exp\left(-\frac{D x}{2\LA x \RA_D}\right) \ ,
  \nonumber \\
  \LA x \RA_D &=& D \LA x \RA_1 = \frac{D k_B T}{2} \ ,
  \label{x_D}
\end{eqnarray}
where $\LA x \RA_D$ represents the average value of kinetic energy 
in $D$ dimensions.
The analogy between the factor $D$ in the argument of the exponential
function in Eq.~(\ref{x_D}) and the analogous factor $n$ 
in Eq.~(\ref{NGibbs}) is to be noticed.
The main difference is that, while $D$ is an integer number by hypothesis,
the parameter $n(\lambda)$ can assume in general any real values
larger than or equal to one.

In Eq.~(\ref{x_D}) temperature appears implicitly
as $T = 2 \LA x \RA_D / k_B D$.
This suggests that also in the closed economy model considered above
the effective temperature of the system should be defined 
as $\LA x \RA/n$, rather than $\LA x \RA$.
This is a natural consequence of the fact that the average value 
of kinetic energy in $D$ dimensions is proportional to $D$,
due to the equipartition theorem, and that an estimate
of the amplitude of thermal fluctuations,
which is independent of its effective dimension, 
can be obtained from the ratio $\LA x \RA_D/D$.

Direct comparison between Eqs.~(\ref{x_D}) and (\ref{NGibbs}) leads to
a formal but exact analogy, 
between money in the closed economy model considered above, with $N$ agents,  
saving propensity $0 \le \lambda \le 1$, and given average money 
$\LA x \RA$, on one hand,
and kinetic energy in an ensemble of $N$ particles
in $D$ dimensions at temperature $T$, on the other, 
if the effective dimension and temperature are defined as
\begin{eqnarray}
  D(\lambda) &=& 2\,n(\lambda) = \frac{2(1 + 2 \lambda)}{1 - \lambda} \ ,
  \nonumber \\
  T(\lambda) &=& \frac{\LA x \RA}{n(\lambda)}
              = \LA x \RA\frac{1 - \lambda}{1 + 2 \lambda} \ ,
\end{eqnarray}
respectively. This equivalence can be qualitatively understood in terms of the
underlying microscopic dynamics by considering
the example of a fluid of interacting particles.
In one dimension, particles undergo head-on collisions, in which they can
exchange the total amount of energy they have.
In an arbitrary (large) number of dimensions, however, 
this is not possible for purely kinematic reasons
and only a fraction of the total energy is actually released or gained 
on average in a collision.
Since the equipartition theorem implies that on average kinetic energy 
is equally shared among the $D$ dimensions,
one can expect that, during a collision, only a fraction $\sim 1/D$ 
of the total energy is exchanged (and that a corresponding 
fraction $\lambda \sim 1-1/D$ is ``saved''). 
This estimate $\sim 1/D$ of the exchanged energy is to be compared 
with the expression for the fraction of exchanged money obtained 
from Eq.~(\ref{n}) using $n=D/2$, namely $1-\lambda = 3/(D/2+2)$,
which was in fact found starting the fitting of the numerical data
from a function prototype of a form similar to $1/D$.

%%%%%%%%%%%%%%%%%%%%%%%%%%%%%%%%%%%%%%%%%%%%%%%%%%%%%%%%%%%%%%%%%
\section{Conclusions}
\label{conclusions}

We have studied a statistical model,
which can be interpreted as a generalization 
of the simple closed economy model, in which a random reassignment 
of the total agent money $x$, involved in the transaction, takes place.
The generalized model is characterized by $N$ agents carrying out
transactions according to a saving criterion,
determined quantitatively through a saving propensity $\lambda > 0$.
Alternatively, it can be considered as representing a gas of $N$ interacting
particles which on average exchange only a fraction of 
their kinetic energy $x$, during a collision.
We have shown the existence of such an analogy,
by empirically obtaining the corresponding analytical solution
$f_n(x)$ for the equilibrium distribution from numerical data.

In both cases the equilibrium distribution can be written as
a Gamma distribution $\gamma_n(\xi)$, 
where the reduced variable is given by $\xi = n x/\LA x \RA_n$
and $\LA x \RA_n$ represents the the average value of $x$,
given by $\LA x \RA_n = n\LA x \RA_1$.
The equivalence is represented by $n = n(\lambda) = 1 + 3\lambda/(1-\lambda)$
being a function of the saving propensity on one hand ($\LA x \RA_1$ is 
the average value for $\lambda=0$),
and by $n=D/2$ being just the half of the number of dimensions,
on the other hand ($\LA x \RA_1$ being in this case the average value
in two dimensions).

The fact that we obtain basically the same equilibrium distribution,
characterizing the kinetic energy a gas of particles,
suggests some general considerations about closed-economy models.
The mechanical analogy illustrated above can be addressed to the fact 
that the system is described statistically by a micro-canonical
ensemble, just as a closed mechanical system, in which the exchanged
quantity is conserved.
Thus, a saving propensity larger than zero or any other change in the
microscopic law can be expected to lead to a different shape of the
equilibrium distribution, as shown in the present work,
in which e.g. the effective number of dimensions and temperature may
be different.
However, the simple fact that the money is conserved implies that one cannot
obtain an arbitrary distribution but, rather, only equilibrium
distributions related directly to the micro-canonical ensemble.
%Due to the Gibbs distribution exponential factor,
%it seems improbable to menage to obtain a power law
%distribution in a closed economy model.
It is suggested that open systems may be characterized by an
equilibrium or stationary state characterized by a type 
of power-law distribution.

%%%%%%%%%%%%%%%%%%%%%%%%%%%%%%%%%%%%%%%%%%%%%%%%%%%%%%%%%%%%%%%%%%

\begin{acknowledgments}
M.P. is grateful to P. Muratore-Ginanneschi for useful discussions.
This work was partially supported by the Academy of Finland,
Research Center for Computational Science and Engineering,
project no. 44897 (Finnish Centre for Excellence Program 2000-2005).
The work at Brookhaven National Laboratory was carried out under Contract No.
DE-AC02-98CH10886, Division of Material Science, U. S. Department of Energy.
\end{acknowledgments}

% Create the reference section using BibTeX:
\bibliography{new}

\end{document}